\documentclass{article}

\begin{document}

\title{An example of double confluent Heun equation: \\
Schr\"odinger equation with supersingular plus Coulomb potential}

\author{Julio Abad and Javier Sesma\footnote{javier@unizar.es}\\  \   \\
Departamento de F\'{\i}sica Te\'{o}rica, \\ Facultad de Ciencias,
\\ 50009 Zaragoza, Spain. \\  \ }

\maketitle

\begin{abstract}
A recently proposed algorithm to obtain global solutions of the double
confluent Heun equation is applied to solve the quantum mechanical problem
of finding the energies and wave functions of a particle bound in a potential
sum of a repulsive supersingular term, $A\,r^{-4}$, plus an attractive Coulombian
one, $-Z\,r^{-1}$. The existence of exact algebraic solutions for certain values of
$A$ is discussed.
\end{abstract}


Supersingular potentials, i. e., potentials presenting at the origin a
singularity of the type $r^{-\alpha}$, with $\alpha > 2$, were firstly
discussed in the context of collision of particles [1--\,6],
the main issue being the adequate
definition of the $S$ matrix. On the other hand, concerned with bound states,
a considerable interest on
repulsive singular potentials added to an attractive regular one was
arisen by the seminal paper of Klauder \cite{klau}, where the today known
as ``Klauder phenomenon" was reported.

As shown by Detwiler and Klauder \cite{detw}, supersingular potentials cannot
be treated by conventional WKB or perturbative methods. In view of this, a great
diversity of approximate methods have been suggested. Extensive lists of articles
applying those methods to supersingular potentials can be found in recent papers by
Saad, Hall and Katatbeh \cite{saa1} and by Liverts and Mandelzweig \cite{live}.

The validity of an approximate method can be inferred from comparison of its
results with those obtained with an exact method. Up to now, direct numerical
integration of the Schr\"odinger equation has been the only ``exact" method to
compare with. Obviously, an algebraic exact method would be preferable.

In order to better elucidate the effect of the supersingular potential on the
energies of a particle bound in an attractive potential, it is much safer
to choose for this one an exactly solvable form. For this reason, the attractive part
of the potential has been taken, in most cases, to be a harmonic oscillator or a
Coulombian well. Nevertheless, even in those simple cases, the resulting
Schr\"odinger equation presents two irregular singularities, one at the origin and
the other at infinity, and solving it requires to deal with the non-trivial issue of
connection between their singular points.

The problem of connecting singular points of a differential equation has been
considered by several authors. Most of them [11--\,14]
refer to a differential equation for which the origin is an ordinary or a
regular singular point and the infinity is an irregular singular
one. As far as we know, only the procedure developed by Naundorf
\cite{naun} becomes applicable also to the case of both the
origin and the infinity being irregular singular points. In a
recent paper \cite{abad}, an algorithm, based on a
modification of the Naundorf's procedure which improves notably its performance,
was presented and applied to find global solutions of the double confluent
Heun equation, a simple case of differential equation with two irregular singular points.
The purpose of this note is to show the usefulness of that algorithm to obtain exact solutions
of the Schr\"odinger equation with supersingular potentials and, therefore, to test the
quality of the different approximate methods proposed to deal with them.
We concentrate on a specific potential, namely an attractive Coulomb potential to which
a singular repulsive term has been added, whose discrete energy levels have been studied
by Aguilera-Navarro {\it et al} \cite{agui} by using variational methods.

The quantum mechanical problem to be solved is that of a particle of mass $m$ and angular
momentum $\hbar l$ bound in a spherically symmetric tridimensional potential whose radial
dependence is given by
\begin{equation}
V(r)=\frac{\hbar^2}{2m}\left(\frac{A\,r_0^2}{r^4}-\frac{Z\,r_0^{-1}}{r}\right),
\qquad A>0, \quad Z>0.\label{uno}
\end{equation}
Here $r_0$ represents an arbitrary length fixing the scale. The factor
$\hbar^2/2m$ in front of the parenthesis guarantees that the parameters
$A$ and $Z$, that represent the intensities of the supersingular and the Coulomb
parts, are dimensionless. The energy of the particle will be denoted by
\[
\frac{\hbar^2}{2mr_0^2}\,E,
\]
where $E$ is also dimensionless. We are here interested on the bound states of the particle.
In other words, we are looking for values of the parameter $E<0$ for which the Schr\"odinger
equation admits normalizable solutions. This requires the wave function to be regular also
at the origin and at infinity, the two singular points of the differential equation.

The changes of independent and dependent variables
\begin{equation}
r/r_0=z \qquad \mbox{and} \qquad R(r)\propto z^{-1/2}\,y(z)  \label{tres}
\end{equation}
turn the Schr\"odinger equation for the radial wave function,
\begin{equation}
\left(-\frac{\hbar^2}{2m}\,\left(\frac{1}{r^2}\frac{d}{dr}r^2\frac{d}{dr} - \frac{l(l+1)}{r^2}\right)
+V(r)\right)R(r)=\frac{\hbar^2}{2mr_0^2}\,E\,R(r),   \label{dos}
\end{equation}
into a particular case of the double confluent Heun equation \cite{schw}
\begin{equation}
D^2\,y(z) + B(z)\,y(z) = 0, \quad D=z\frac{d}{dz}, \quad B(z)=\sum_{p=-2}^2B_p\,z^p,  \label{cuatro}
\end{equation}
with parameters
\begin{equation}
B_{-2}=-A,\quad B_{-1}=0,\quad B_0=-l(l\!+\!1)-1/4,\quad B_1=Z, \quad B_2=E.  \label{cinco}
\end{equation}
The algorithm described in Ref. \cite{abad} deals with the function
\begin{equation}
w(z)=z^{1/2}y(z),    \label{seis}
\end{equation}
which turns out to be proportional to the reduced radial wave function,
\begin{equation}
w(z)\propto rR(r),  \label{siete}
\end{equation}
and satisfies the differential equation
\begin{equation}
z^2\,\frac{d^2w}{dz^2} + \left(-\frac{A}{z^2}-l(l+1)+Z\,z+E\,z^2\right)w=0. \label{ocho}
\end{equation}
Now it is immediate to use our method to determine the eigenvalues $E$ and the
(reduced) eigenfunctions $w$.

The differential equation (\ref{ocho}) admits two {\em Floquet} solutions whose
Laurent expansions are
\begin{equation}
w_j=z^{\nu_j}\sum_{n=-\infty}^{\infty} c_{n,j}\,z^n,  \qquad j=1,2, \label{nueve}
\end{equation}
the coefficients $c_{n,j}$ obeying a fourth order recurrence relation,
\[
-A\,c_{n+2,j}+\left((n+\nu_j)(n-1+\nu_j)-l(l+1)\right)c_{n,j}+Z\,c_{n-1,j}+E\,c_{n-2,j}=0.
\]
The requirement of those expansions to be convergent, i. e., the condition
\begin{equation}
\sum_{n=-\infty}^{\infty} |c_{n,j}|^2<\infty,  \label{diez}
\end{equation}
fixes, for given values of the parameters $A$, $l$, $Z$, and $E$, each one
of the indices $\nu_j$, except for addition of an arbitrary integer accompanied of a
relabeling of the coefficients $c_{n,j}$. Any solution of (\ref{ocho}) can then be
written as a linear combination
\begin{equation}
w = \zeta_1\,w_1 + \zeta_2\,w_2     \label{duno}
\end{equation}
of the two Floquet solutions. To be acceptable from the physical point of view,
the behaviour of that combination at the singular points must be
\begin{eqnarray}
w(z) & \propto & \exp (-\alpha z)\,z^{\mu}\left(1+O(z^{-1})\right), \quad \mbox{as}
\quad z\to\infty,  \label{ddos} \\
w(z) & \propto  & \exp (-\beta z^{-1})\, z\, \left(1+O(z)\right), \;\;\quad \mbox{as}
\quad z\to 0,   \label{dtres}
\end{eqnarray}
with exponents
\[
\alpha=\sqrt{-E},\qquad \mu=\frac{Z}{2\sqrt{-E}},\qquad \beta=\sqrt{A}.
\]
Although the combination (\ref{duno}) contains two arbitrary coefficients, $\zeta_1$
and $\zeta_2$, the homogeneity of the differential equation reduces to only one the
number of effective degrees of freedom of that combination. Then, for given $A$, $l$
and $Z$, both conditions (\ref{ddos}) and (\ref{dtres}) are satisfied  only for
the values of $E$ corresponding to bound states.

We have applied our algorithm \cite{abad} to the determination of the energy of
the lowest states bound in the potential (\ref{uno}) for the set of values of
the parameter $A$ considered by Aguilera-Navarro {\it et al} \cite{agui}. The
Coulombian parameter $Z$ has been taken, following those authors, equal to 1.
The results corresponding to any other value of $Z>0$, for instance $\hat{Z}$,
can be inferred from those shown in Table 1 by noting that a trivial change of
scale transforms Eq. (\ref{ocho}), with $Z=1$,  into
\[
z^2\,\frac{d^2w}{dz^2} + \left(-\frac{\hat{A}}{z^2}-l(l+1)+\hat{Z}\,z+\hat{E}\,z^2\right)w=0,
\]
with $\hat{A}=\hat{Z}^{-2}A$ and $\hat{E}=\hat{Z}^2E$.
\begin{table}
\caption{Energies $E_{n,l}$ of the ground ($n=0$), first excited ($n=1$),
and second excited ($n=2$) states of a particle of angular momentum  $l=0$, $1$ or $2$
in the potential given by Eq. (\ref{uno}) with $Z=1$ and different values of $A$.
The energies are given in units $\hbar^2/2mr_0^2$. In absence ($A=0$) of a
supersingular term, the energies are the Coulombian ones, $E_{n,l}=1/4(n+l+1)^2$.}
\begin{center}
\begin{tabular}{ccccc}
\hline
$l$ & $A$ & $E_{0,l}$ & $E_{1,l}$ & $E_{2,l}$ \\
\hline
\ & \ & \ & \ & \   \\
$0$ & $0.0001$ & $-0.245429530$ & $-0.061924800$ & $-0.027606954$ \\
\  & $0.01$ & $-0.218348228$ & $-0.058346200$ & $-0.026526103$ \\
\  & $1$ & $-0.139037013$ & $-0.045670342$ & $-0.022416709$ \\
\  & $25$ & $-0.077060194$ & $-0.031979568$ & $-0.017358131$ \\
\  & $100$ & $-0.056304895$ & $-0.026031554$ & $-0.014892512$ \\
\ & \ & \ & \ & \  \\
$1$ & $0.0001$ & $-0.062499741$ & $-0.027777693$ & $-0.015624935$ \\
\ & $0.01$ & $-0.062475231$ & $-0.027769650$ & $-0.015621471$ \\
\ & $1$ & $-0.060877715$ & $-0.027256429$ & $-0.015398665$ \\
\ & $25$ & $-0.050470052$ & $-0.023926758$ & $-0.013937057$ \\
\ & $100$ & $-0.042059190$ & $-0.021077780$ & $-0.012640912$ \\
\ & \ & \ & \ & \  \\
$2$ & $0.0001$ & $-0.027777774$ & $-0.015624998$ & $-0.00999998$ \\
\ & $0.01$ & $-0.027777435$ & $-0.015624841$ & $-0.009999914$ \\
\ & $1$ & $-0.027743922$ & $-0.015608960$ & $-0.009991373$ \\
\ & $25$ & $-0.027076025$ & $-0.015299319$ & $-0.009825968$ \\
\ & $100$ & $-0.025711754$ & $-0.014683511$ & $-0.009499543$ \\
\hline
\end{tabular}
\end{center}
\end{table}

Besides the values of the energy, our algorithm provides with the indices
$\nu_j$ and the coefficients $c_{n,j}$ of the Floquet solutions and with the
values of the coefficients $\zeta_1$ and $\zeta_2$ that make the linear
combination (\ref{duno}) to be well behaved at the singular points, i. e.,
the (to be normalized) solution of the physical problem. In the neighbourhood
of the singular points, the wave function should be evaluated by making use of
its asymptotic expansions
\begin{eqnarray}
w(z) & \sim & \,\exp (-\alpha z)\,z^{\mu}\sum_{m=0}^\infty a_m\,z^{-m}, \quad \mbox{as}
\quad z\to\infty,  \label{dcuatro} \\
w(z) & \sim  & \,\exp (-\beta z^{-1})\, z \sum_{m=0}^\infty b_m\,z^m, \;\quad \mbox{as}
\quad z\to 0,   \label{dcinco}
\end{eqnarray}
where the first coefficients $a_0$ and $b_0$ are given by our algorithm and the rest of them
are obtained from the recurrence relations
\begin{eqnarray}
-2\,\alpha\, m\, a_m & = & \left((m-\mu)(m-1-\mu)-l(l+1)\right)a_{m-1} - A\, a_{m-3}  \label{dseis} \\
-2\,\beta\, m\, b_m & =  & \left(m(m-1)-l(l+1)\right)b_{m-1} + Z \,b_{m-2} + E\, b_{m-3},   \label{dsiete}
\end{eqnarray}
As an example, let us consider the $l=0$ ground state in a potential of intensity $A=10$. Our algorithm
gives for the energy the value $E=-0.093111277969$. For the indices
of the Floquet solutions we obtain pure imaginary values, $\nu_1=-\nu_2 =0.918988880508\,i$. So,
besides of opposite, they turn to be complex conjugate to each other. In this case, since the
parameters of the potential are real, the coefficients $c_{n,j}$ of the Laurent expansions
(\ref{nueve}) of the two Floquet solutions are complex conjugate, $c_{n,2}=\overline{c_{n,1}}$,
and $w_2(z)=\overline{w_1(z)}$ for real $z$. To give an idea of the magnitude of the coefficients
of the Laurent expansion, we report in Table 2 some of those of $w_1$ normalized
in such a way that $c_{n,0}=1$. A physically acceptable solution results if we take a linear combination
as in (\ref{duno}) with
\[
\zeta_1 =  -0.059000052486 + 0.998257979586\,i, \qquad \zeta_2 = -\,\overline{\zeta_1}.
\]
The coefficients $a_0$ and $b_0$ of the asymptotic expansions (\ref{dcuatro}) and (\ref{dcinco}) are then
\[
a_0 = -2.5451305418\, i, \qquad b_0 = -4.1825205880\, i.
\]
\begin{table}
\caption{Real and imaginary parts of the coefficients $c_{n,1}$ of the Laurent expansion, Eq. (\ref{nueve}),
of one of the Floquet solutions, $w_1$, of the differential equation (\ref{ocho}) with $A=10$, $l=0$, $Z=1$, and
$E=-0.093111277969$. The normalization of $w_1$ has been arbitrarily taken such that $c_{0,1}=1$.
The index (see the text) turns to be $\nu_1=0.918988880508\,i$. The other Floquet
solution, $w_2$, has coefficients $c_{n,2}=\overline{c_{n,1}}$ and index $\nu_2=\overline{\nu_1}$.}
\begin{center}
\begin{tabular}{rrr}
\hline
$n$ & $\Re c_{n,1}\qquad\qquad$ & $\Im c_{n,1}\qquad\qquad$ \\
\hline
$-20$ & $-0.121469106359$E$-09$ & $0.131600927337$E$-09$ \\
$-19$ & $-0.597147461925$E$-09$ & $-0.103014061043$E$-08$ \\
$-18$ & $-0.459666139538$E$-08$ & $0.597145217189$E$-08$ \\
$-17$ & $-0.263452992860$E$-07$ & $-0.369047564493$E$-07$ \\
$-16$ & $-0.136571713612$E$-06$ & $0.219244950133$E$-06$ \\
$-15$ & $-0.923098009638$E$-06$ & $-0.104082342414$E$-05$ \\
$-14$ & $-0.304091218966$E$-05$ & $0.635537646613$E$-05$ \\
$-13$ & $-0.250549707024$E$-04$ & $-0.222398101994$E$-04$ \\
$-12$ & $-0.467558678970$E$-04$ & $0.140924281725$E$-03$ \\
$-11$ & $-0.509362942365$E$-03$ & $-0.340072927974$E$-03$ \\
$-10$ & $-0.404150379497$E$-03$ & $0.229165431769$E$-02$ \\
$-9$ & $-0.740381881993$E$-02$ & $-0.336931518078$E$-02$ \\
$-8$ & $-0.394128245178$E$-04$ & $0.257592989152$E$-01$ \\
$-7$ & $-0.719278491963$E$-01$ & $-0.168793007782$E$-01$ \\
$-6$ & $0.392262032139$E$-01$ & $0.182994779417$E$+00$ \\
$-5$ & $-0.419924194846$E$+00$ & $0.866008349063$E$-02$ \\
$-4$ & $0.372866041301$E$+00$ & $0.704332666002$E$+00$ \\
$-3$ & $-0.121096156077$E$+01$ & $0.468201746853$E$+00$ \\
$-2$ & $0.125443087271$E$+01$ & $0.103994993070$E$+01$ \\
$-1$ & $-0.100849614575$E$+01$ & $0.137165533736$E$+01$ \\
$0$ & $0.100000000000$E$+01$ & $0.000000000000$E$+00$ \\
$1$ & $0.398351667070$E$+00$ & $0.536163740419$E$+00$ \\
$2$ & $-0.196983836993$E$+00$ & $0.355835389786$E$-01$ \\
$3$ & $0.264749708403$E$-01$ & $-0.214447855751$E$-01$ \\
$4$ & $-0.204690743398$E$-02$ & $0.342012088434$E$-02$ \\
$5$ & $0.953365922946$E$-04$ & $-0.324605520121$E$-03$ \\
$6$ & $-0.191766489961$E$-05$ & $0.227409233686$E$-04$ \\
$7$ & $-0.102980463212$E$-06$ & $-0.125761848401$E$-05$ \\
$8$ & $0.131145372215$E$-07$ & $0.579282461546$E$-07$ \\
$9$ & $-0.819888649623$E$-09$ & $-0.228010314464$E$-08$ \\
$10$ & $0.382880819640$E$-10$ & $0.785818186305$E$-10$ \\
$11$ & $-0.147533767446$E$-11$ & $-0.240418109468$E$-11$ \\
$12$ & $0.491035084921$E$-13$ & $0.662078837429$E$-13$ \\
$13$ & $-0.144702107807$E$-14$ & $-0.165529043300$E$-14$ \\
$14$ & $0.384193021031$E$-16$ & $0.379059006228$E$-16$ \\
$15$ & $-0.929783698355$E$-18$ & $-0.799671152250$E$-18$ \\
$16$ & $0.207085395184$E$-19$ & $0.156351350772$E$-19$ \\
$17$ & $-0.427466376462$E$-21$ & $-0.284451924206$E$-21$ \\
$18$ & $0.822932028905$E$-23$ & $0.483549589235$E$-23$ \\
$19$ & $-0.148466844882$E$-24$ & $-0.770120556643$E$-25$ \\
$20$ & $0.252142755003$E$-26$ & $0.115225564234$E$-26$ \\
\hline
\end{tabular}
\end{center}
\end{table}

In the above discussed example, the indices $\nu_j$ of the Floquet solutions
happen to be pure imaginary. This is not always the case. We report, in Table 3,
those indices for the most tightly bound states, with different angular momenta,
in potentials corresponding to several values of $A$.
\begin{table}
\caption{Indices $\nu_1$ of one of the Floquet solutions of Eq. (\ref{ocho})
with $Z=1$, the shown values of $A$ and $l$, and the energy $E$ of the most tightly
bound state. The indices of the other Floquet solution are the opposite, $\nu_2=-\nu_1$.
Notice that $\nu_2=\overline{\nu_1}$ if $\nu_1$ is pure imaginary. In the cases
of complex $\nu_1$, since $\Re \nu_1=1/2$, the ambiguity in the definition of the
indices allows one to take $\nu_2=-\nu_1+1=\overline{\nu_1}$. This choice makes the set
of coefficients of the Laurent expansion of one of the Floquet solutions to be the complex
conjugate of the set of the other one.}
\begin{center}
\begin{tabular}{ccrr}
\hline
\hspace{0.5cm}$l$\hspace{0.5cm} & \hspace{0.5cm}$A$\hspace{0.5cm} & $\nu$\hspace{1.0cm} & $E$\hspace{1.0cm} \\
\hline
$0$ & $0.5$ & $0.487816983037\,i$ & \hspace{1.0cm}$-0.1533318066$ \\
\  & $5$ & $0.825876276087\,i$ & $-0.1062948831$ \\
\  & $65$ & $0.5 + 0.556566003844\,i$ & $-0.0622769642$ \\
\  & \  & \  & \  \\
$1$  & $0.5$ & $0.020895739155$ & $-0.0615761515$ \\
\  & $5$ & $0.229880095066$ & $-0.0574288855$ \\
\  & $65$ & $0.5 + 0.898656568840\,i$ & $-0.0448293862$ \\
\  & \  & \  & \  \\
$2$  & $0.5$ & $0.000919514453$ & $-0.0277607452$ \\
\  & $5$ & $0.008595680010$ & $-0.0276154597$ \\
\  & $65$ & $0.108838373090$ & $-0.0262524928$ \\
\hline
\end{tabular}
\end{center}
\end{table}
It is interesting to notice that, given $l$, the indices $\nu_j$ vanish for
certain values of $A$. In these cases, the wave function could be written in the closed form
\begin{equation}
w(z) = \exp (-\alpha z-\beta z^{-1})\,z\,v(z),  \label{docho}
\end{equation}
where $v(z)$ represents a series of integer powers of $z$, let us say
\begin{equation}
v(z) =\sum_{j=0}^\infty\xi_j\,z^j.  \label{dnueve}
\end{equation}
The question arises if this series could in fact turn to
be a polynomial for any values of $A$, that is, if $w(z)$ is a {\em quasi-polynomial} solution,
in the terminology of Schmidt and Wolf \cite{schw}. Comparison of
(\ref{docho}) with (\ref{dcuatro}) and (\ref{dcinco}) makes evident that $\mu$ should be an integer,
for instance $p$. Accordingly, the energy parameter would be
\begin{equation}
E=-\,\frac{Z}{4p^2}  \label{veinte}
\end{equation}
whereas the parameter $A$ should be such that the coefficients $\xi_j$ in (\ref{dnueve}) vanish for $j\geq p$.
Obviously, the $\xi_j$ are related to the coefficients $a_m$ of the asymptotic expansion (\ref{dcuatro}) by
\begin{equation}
\exp(-\beta z^{-1})\sum_{j=0}^{p-1}\xi_j\,z^j = z^{p-1}\,\sum_{m=0}^\infty a_m\,z^{-m}, \label{vuno}
\end{equation}
and to the $b_m$ of (\ref{dcinco}) by
\begin{equation}
\exp(-\alpha z)\sum_{j=0}^{p-1}\xi_j\,z^j = \sum_{m=0}^\infty b_m\,z^{m}. \label{vdos}
\end{equation}
The relations
\begin{equation}
\xi_{p-1-k} = a_k - \sum_{s=1}^k \frac{(-\beta)^s}{s!}\,\xi_{p-1-k+s}, \qquad k=0, 1, \ldots, p-1,
\label{vtres}
\end{equation}
following immediately from Eq. (\ref{vuno}), allow one to write explicit expressions of the $\xi_j$
in terms of $\beta$. Substitution of these expressions in the relation
\begin{equation}
0 = a_p - \sum_{s=1}^p \frac{(-\beta)^s}{s!}\,\xi_{s-1},  \label{vcuatro}
\end{equation}
following also from (\ref{vuno}), produces a polynomial equation for $\beta$ whose solution
provides the values of $A=\beta^2$ for which the wave function is given by (\ref{docho}),
$v(z)$ being a polynomial of degree $p-1$. Table 4 shows several sets of values of $A$ and $l$ for which the energy
$E$ and the wave function $w$ of a (ground or excited) state admit such particularly simple expressions.
Solutions of that kind, i. e., quasi-polynomial solutions, are especially interesting to test the accuracy
of the different approximate methods.
\begin{table}
\caption{Sets of values of the parameters $A\neq 0$, $l$, and $E$ for which Eq. (\ref{ocho}) with $Z=1$
admits a particularly simple solution of the form (\ref{docho}), $v(z)$ being a polynomial. The list
could be continued, the values of $E$ being given by $E=-1/4p^2$, with integer $p$.}
\begin{center}
\begin{tabular}{ccc}
\hline
\hspace{0.6cm}$E$\hspace{0.6cm} & \hspace{0.6cm} $l$\ \hspace{0.6cm} & $\sqrt{A}$ \\
\hline
\  & \ &  \\
$-1/16$ & $0$ & $8$ \\
\  & \ &  \\
\  & \ &  \\
$-1/36$ & $0$ & $6\left(4+\sqrt{7}\right)$ \\
\  & \  & $6\left(4-\sqrt{7}\right)$ \\
\  & \ &  \\
\  & $1$ & $30$ \\
\  & \ &  \\
\  & \ &  \\
$-1/64$ & $0$ &  $(16/3)\left(10+\Re\left\{\left[2(436+i\,9\sqrt{4434})\right]^{1/3}\right\}\right)$ \\
\  & \  & $(8/3)\left(20-\Re\left\{\left(1+i\,\sqrt{3}\right)\left[2(436+i\,9\sqrt{4434})\right]^{1/3}\right\}\right)$ \\
\  & \  & $(8/3)\left(20-\Re\left\{\left(1-i\,\sqrt{3}\right)\left[2(436+i\,9\sqrt{4434})\right]^{1/3}\right\}\right)$ \\
\  & \ &  \\
\  & $1$ & $8\left(8+3\sqrt{2}\right)$  \\
\  & \  & $8\left(8-3\sqrt{2}\right)$  \\
\  & \ &  \\
\  & $2$ & $8\left(4+\sqrt{22}\right)$  \\
\  & \ &  \\
\  & \ &  \\
$-1/100$ & $0$ & $5\left(20+\epsilon+\left[310-\epsilon^2+1080/\epsilon\right]^{1/2}\right)$ \\
\  & \  & $5\left(20+\epsilon-\left[310-\epsilon^2+1080/\epsilon\right]^{1/2}\right)$ \\
\  & \  & $5\left(20-\epsilon+\left[310-\epsilon^2-1080/\epsilon\right]^{1/2}\right)$ \\
\  & \  & $5\left(20-\epsilon-\left[310-\epsilon^2-1080/\epsilon\right]^{1/2}\right)$ \\
\  & \  & $\left(\epsilon\equiv\left[\left(310+2\Re\left\{\left[25\left(231709+i\,108\sqrt{1810371}\right)\right]^{1/3}\right\}\right)/3\right]^{1/2}\right)$ \\
\  & \ &  \\
\  & $1$ & $10\left(16+\sqrt{37}\right)$ \\
\  & \  & $10\left(16-\sqrt{37}\right)$ \\
\  & \  & $30$ \\
\  & \ &  \\
\  & $2$ & $(10/3)\left(25+2\Re\left\{\left[4\left(439+i\,9\sqrt{14559}\right)\right]^{1/3}\right\}\right)$ \\
\  & \  & $(10/3)\left(25-\Re\left\{\left(1+i\,\sqrt{3}\right)\left[4\left(439+i\,9\sqrt{14559}\right)\right]^{1/3}\right\}\right)$ \\
\  & \ &  \\
\  & $3$ & $(10/3)\left(10+\left[3142-9\sqrt{15519}\right]^{1/3}+\left[3142+9\sqrt{15519}\right]^{1/3}\right)$ \\
\hline
\end{tabular}
\end{center}
\end{table}
The same results
are obtained starting from Eq. (\ref{vdos}) instead of (\ref{vuno}). The relations defining the $\xi_j$
are then
\begin{equation}
\xi_{k} = b_k - \sum_{s=1}^k \frac{(-\alpha)^s}{s!}\,\xi_{k-s}, \qquad k=0, 1, \ldots, p-1,
\label{vcinco}
\end{equation}
and the equation whose fulfilment determines $\beta$ turns out to be
\begin{equation}
0 = b_p - \sum_{s=1}^p \frac{(-\alpha)^s}{s!}\,\xi_{p-s}.  \label{vseis}
\end{equation}
Notice that this equation expresses the fact that $\xi_p$, defined as in (\ref{vcinco}) with
$k=p$, vanishes. On the other hand, substitution of (\ref{docho}) and  (\ref{dnueve}) in the differential equation
(\ref{ocho}) reveals the recurrence relation
\begin{equation}
-2\beta j\, \xi_j = \left(j(j-1)-l(l+1)-2\alpha\beta\right)\,\xi_{j-1}+\left(Z-2\alpha(j-1)\right)\,\xi_{j-2}.
\label{vsiete}
\end{equation}
Therefore, since $2\alpha=Z/p$, all $\xi_j$ with $j>p$ vanish if $\xi_p=0$. By the way, the last recurrence relation
provides a third procedure to determine the values of $A$ for which a quasi-polynomial solution
exists: start with $\xi_0$ and $\xi_1$ as given by (\ref{vcinco}), use the recurrence (\ref{vsiete})
to write $\xi_p$ in terms of $\beta$, and then solve the equation $\xi_p=0$.

The results shown above have been obtained by using double precision FORTRAN codes.
Additional digits in the results would require higher precision arithmetic. This is especially necessary
in the case of excited states, due to the progressive decreasing of the energy gap between consecutive levels.

\bigskip
Financial support from Comisi\'{o}n Inter\-mi\-nis\-te\-rial de
Ciencia y Tecnolog\'{\i}a and of Diputaci\'on General de Arag\'on
is acknowledged.


\begin{thebibliography}{99}

\bibitem{fubi} {S. Fubini  and R. Stroffolini, Nuovo Cimento
{\bf 37} (1965) 1812.}%

\bibitem{fran} {W. M. Frank, D. J. Land and R. M. Spector, Rev. Mod. Phys.
{\bf 43} (1971) 36.}%

\bibitem{baet} {M. L. Baeteman, Nucl. Phys. {\bf A235} (1974) 249.}%

\bibitem{bueh} {W. B\"uhring, J. Math. Phys. {\bf 15} (1974) 1451.}%

\bibitem{robi} {D. W. Robinson, Ann. Ins. Henri Poincar\'e {\bf 21} (1974) 185.}%

\bibitem{aly} {H. H. Aly, H. J. W. M\"uller-Kirsten and N. Vahedi-Faridi,
 J. Math. Phys. {\bf 16} (1975) 961.}%

\bibitem{klau} {J. R. Klauder, Science {\bf 199} (1978) 735.}%

\bibitem{detw} {L. C. Detwiler and J. R. Klauder, Phys Rev. D
{\bf 11} (1975) 1436.}%

\bibitem{saa1} {N. Saad, R. L. Hall and Q. D. Katatbeh, J.
Math. Phys. {\bf 46} (2005) 022104.}%

\bibitem{live} {E. Z. Liverts and V. B. Mandelzweig, Ann.
Phys. (N. Y.) {\bf 322} (2007) 2211.}%

\bibitem{kohn} {M. Kohno, Hiroshima Math. J. {\bf 4} (1974) 293.}%

\bibitem{sch1} {R. Sch\"afke and D. Schmidt, SIAM J. Math. Anal. {\bf 11}
(1980) 848.}%

\bibitem{sch2} {R. Sch\"afke, SIAM J. Math. Anal. {\bf 11} (1980) 863.}%

\bibitem{sch3} {R. Sch\"afke, SIAM J. Math. Anal. {\bf 15} (1984) 253.}%

\bibitem{naun} {F. Naundorf, SIAM J. Math. Anal. {\bf 7} (1976) 157.}%

\bibitem{abad} {J. Abad, F. J. G\'omez and J. Sesma, Numer. Algor. {\bf 49} (2008) 33.}%

\bibitem{agui} {V. C. Aguilera-Navarro, E. Ley-Koo
and S. Mateos-Cort\'es, Int. J. Theoret. Phys. {\bf 40} (2001) 1809.}%

\bibitem{schw} {D. Schmidt and G. Wolf, Double confluent Heun equation. In: A.
Ronveaux (ed.), Heun's Differential Equations, Oxford
University Press, Oxford (1995), pp. 129--188.}%

\end{thebibliography}
\end{document}